\documentclass[pre,twocolumn,showpacs,floatfix]{revtex4-1}

\usepackage{amssymb,amsmath}

\usepackage{epsfig}

\usepackage{graphicx}
\usepackage{color}

\newcommand{\dd}{\text{d}}

\begin{document}
\title{Segregation of an intruder in a heated granular dense gas}
\author{Vicente Garz\'{o}}
\email{vicenteg@unex.es}
\homepage{URL: http://www.unex.es/eweb/fisteor/vicente/}
\affiliation{Departamento de F\'{\i}sica, Universidad de Extremadura, E-06071 Badajoz, Spain}
\author{Francisco Vega Reyes}
\email{fvega@unex.es}
\affiliation{Departamento de F\'{\i}sica, Universidad de Extremadura, E-06071 Badajoz, Spain}

\begin{abstract}

A recent segregation criterion [V. Garz\'o, Phys. Rev. E \textbf{78}, 020301(R) (2008)] based on the thermal diffusion factor $\Lambda$ of an intruder in a heated granular gas described by the inelastic Enskog equation is revisited. The sign of $\Lambda$ provides a criterion for the transition between the Brazil-nut effect (BNE) and the reverse Brazil-nut effect (RBNE). The present theory incorporates two extra ingredients not accounted for by the previous theoretical attempt. First, the theory is based upon the second Sonine approximation to the transport coefficients of the mass flux of intruder. Second, the dependence of the temperature ratio (intruder temperature over that of the host granular gas) on the solid volume fraction is taken into account in the first and second Sonine approximations. In order to check the accuracy of the Sonine approximation considered, the Enskog equation is also numerically solved by means of the direct simulation Monte Carlo (DSMC) method to get the kinetic diffusion coefficient $D_0$. The comparison between theory and simulation shows that the second Sonine approximation to $D_0$ yields an improvement over the first Sonine approximation when the intruder is lighter than the gas particles in the range of large inelasticity. With respect to the form of the phase diagrams for the BNE/RBNE transition, the kinetic theory results for the factor $\Lambda$ indicate that while the form of these diagrams depends sensitively on the order of the Sonine approximation considered  when gravity is absent, no significant differences between both Sonine solutions appear in the opposite limit (gravity dominates the thermal gradient). In the former case (no gravity), the first Sonine approximation overestimates both the RBNE region and the influence of dissipation on thermal diffusion segregation.
\end{abstract}

\draft 
\date{\today}
\maketitle

\section{Introduction}
\label{sec1}

The understanding of the physical mechanisms involved in the segregation of an intruder in a granular fluid is perhaps one of the most important open challenges of granular flow research. This problem has spawned a number of important experimental, computational, and theoretical works in the field of granular media \cite{K04}. Among the different mechanisms proposed to describe (size) segregation, thermal diffusion becomes the most relevant if the system fulfills the conditions of a granular gas. In this case, kinetic theory, properly modified to account for the inelasticity of collisions, has proven to be a reliable tool to analyze the dynamics of the intruder.

Thermal diffusion (or thermophoresis in its single-particle manifestation \cite{GR83}) is the transport of matter due to the presence of a thermal gradient. As a result of the motion of the components of the mixture, a steady state can be reached in which the separating effect arising from thermal diffusion is balanced by the remixing effect of ordinary diffusion. As a consequence, partial segregation is observed and described by the so-called thermal diffusion factor $\Lambda$. While this phenomenon has been widely studied in ordinary gases and liquids \cite{TDF}, much less is known in the case of granular mixtures. It must be noted that in the latter case thermal diffusion can appear in vibrated systems even in the absence of an external imposed
temperature gradient, as a consequence of inelasticity. In this case (energy supplied by vertical walls), the mean kinetic energy of the grains decays away from the vibrating plate giving rise to a (granular) temperature gradient.

In a non-convecting steady state with gradients only along the vertical direction ($z$ axis), $\Lambda$ is defined by the relation \cite{KCM87}
\begin{equation}
\label{1} -\Lambda\frac{\partial \ln T}{\partial z} =\frac{\partial}{\partial z}\ln
\left(\frac{n_0}{n}\right),
\end{equation}
where $T$ is the (granular) temperature, and $n_0$ and $n$ are the number densities of the intruder (or tracer particles in a binary mixture) and gas particles, respectively. If one assumes that gravity ${\bf g}$ and the temperature gradient point in parallel directions (i.e., the bottom plate is hotter than the top plate, $\partial_z \ln T<0$), then the intruder rises with respect to the gas particles if $\Lambda>0$ while the opposite happens  if $\Lambda<0$. When the intruder is larger than the gas particles, the former situation is referred to as the Brazil-nut effect (BNE) while the latter is called the reverse Brazil-nut effect (RBNE). Therefore, the sign of the thermal diffusion factor provides a criterion for the transition between the BNE and the RBNE by varying the parameters of the system (intruder plus granular gas).

A segregation criterion based on the knowledge of the factor $\Lambda$ has been recently obtained \cite{G08} from a solution of the (inelastic) Enskog kinetic equation that applies to first order in the spatial gradients (Navier-Stokes order). In contrast to previous theoretical attempts \cite{JY02,TAH03}, the approach is not limited to near elastic particles and takes into account the combined effect of thermal gradient and gravity. On the other hand, the results reported in Ref.\ \cite{G08} are based on two simplifying assumptions. First, although not explicitly stated, they were obtained by neglecting the dependence of the temperature ratio $\gamma\equiv T_0/T$ between the intruder and the gas particles on the volume fraction $\phi$. This assumption might be more questionable as the gas becomes denser (see for instance, Fig.\ \ref{fign1} below). Second, the results of Ref.\ \cite{G08} were derived by using the first Sonine approximation to estimate the transport coefficients associated with the mass flux of intruder. However, recent results \cite{GF09} obtained in the tracer limit
for an \emph{undriven} granular gas clearly show that the accuracy of the first Sonine solution can
be worsen for small values of the coefficients of restitution and/or disparate values of the mass
and size ratios \cite{notebis}. The question arises then as to whether, and if so to what extent, the conclusions
drawn from Ref.\ \cite{G08} may be altered when the above two new ingredients (density dependence of the temperature ratio and second Sonine correction) are accounted for in the theory. In this paper, we address this question by determining the thermal diffusion factor from the first and second Sonine approximations.

It is important to remark that the density dependence of the temperature ratio affects to both Sonine approximations. In this sense, what is new in the present paper is not only that the calculations are carried out to higher orders, but also that the contributions coming from the term $\partial_\phi \gamma$ are completely accounted for in the first and second Sonine solutions. This fundamental dependence was not considered in the previous works on thermal diffusion for driven \cite{G08,G09} and undriven \cite{GF09} granular gases. As the results show, both improvements (density dependence of temperature ratio and second Sonine correction) have a major impact on the physics of the system for large mechanical differences between host and intruder particles.

On the other hand, it must be pointed out that the segregation criterion derived here considers the temperature gradient as an input and not created by the inelasticity of grains. Thus, if the temperature gradient has a given form, then the segregation criterion has a resultant form. However, it is known \cite{BRM01,BT02bis,BRM05,SGNT06} for vertically vibrated granular systems that after the decrease in the value of the granular temperature as a function of height above the floor, the temperature profile possesses a minimum above which the temperature \emph{increases} as a function of height. Thus, given that $\partial_zT<0$ in Eq.\ \eqref{1}, our segregation criterion can be useful to analyze situations where the system is sufficiently small (shallow layers) so that the minimum in the temperature profile is not reached or is very close to the top of the sample.

The plan of the paper is as follows. First, the thermal diffusion factor $\Lambda$ is  evaluated in Sec.\ \ref{sec2} by using a hydrodynamic description. This factor is expressed in terms of the pressure $p$ of the gas, the transport coefficients $D_0$, $D$, and $D^T$ associated with the mass flux of intruder, and the (reduced) gravity $g^*$. As for ordinary gases \cite{CC70}, the above transport coefficients obey a set of coupled linear integral equations that can be approximately solved by using a Sonine polynomial expansion. These coefficients are determined in Sec.\ \ref{sec3} by retaining terms up to the second Sonine approximation. Some technical details of the calculations are relegated to the Appendix \ref{appA}. To assess the reliability of the first and second Sonine approximations, we compare in Sec.\ \ref{sec4} the kinetic theory predictions for the kinetic diffusion coefficient $D_0$ with numerical simulations of the Enskog equation by using the DSMC method \cite{B94,Chema}. As in the undriven case \cite{GF09}, the coefficient $D_0$ is computed from the mean-square displacement of intruders immersed in a \emph{heated} dense granular gas. The knowledge of $D_0$, $D$, and $D^T$ allows one to express the factor $\Lambda$ in terms of the parameter space of the system: the mass ($m_0/m$) and diameter ($\sigma_0/\sigma$) ratios, the solid volume fraction $\phi$ and the coefficients of restitution $\alpha$ and $\alpha_0$ characterizing dissipation of gas-gas and intruder-gas collisions, respectively. In order to assess the impact of the above parameters on segregation, in Sec.\ \ref{sec5} the form of the BNE/RBNE phase diagrams in the $\left\{\sigma_0/\sigma, m_0/m \right\}$-plane is investigated by varying the parameters of the system in the case of hard spheres with a common coefficient of restitution ($\alpha=\alpha_0$). Two different limit situations are mainly analyzed: (i) absence of gravity  and (ii) thermalized systems (gravity dominates the temperature gradient). The paper is closed in Sec.\ \ref{sec6} with a discussion of the results.

\section{Thermal diffusion of an intruder. Hydrodynamic description}
\label{sec2}

Let us consider a binary mixture of inelastic hard disks ($d=2$) or spheres ($d=3$) where the
concentration of one of the species (of mass $m_0$ and diameter $\sigma_0$) is very small compared to that of the other (excess component of mass $m$ and diameter $\sigma<\sigma_0$). The inelasticity of collisions among gas-gas and intruder-gas is accounted for by (constant) coefficients of normal restitution $\alpha$ and $\alpha_0$, respectively. The system (gas plus intruder) is in presence of the gravitational field $\textbf{g}=-g \widehat{\textbf{e}}_z$, where $g$ is a positive constant and $\widehat{\textbf{e}}_z$ is the unit vector in the positive direction of the $z$ axis.

As mentioned in the Introduction, we consider an inhomogeneous nonconvecting \emph{steady} state with only gradients in the $z$ direction. Since no shearing flows are present, then the pressure tensor $P_{ij}$ of the gas is diagonal, namely, $P_{ij}=p\delta_{ij}$, where $p$ is the hydrostatic pressure. In this case, the momentum balance equation for the gas becomes \cite{G09}
\begin{equation}
\label{2.1}
\frac{\partial p}{\partial z}=-\rho g,
\end{equation}
where $\rho=mn$ is the mass density of the gas particles. In the context of the Enskog equation, the hydrostatic pressure $p$ is given by \cite{GDH07},
\begin{equation}
\label{n1}
p=nT[1+2^{d-2}\chi\phi(1+\alpha],
\end{equation}
where $\chi(\phi)$ is the contact value of the pair correlation function for the granular gas and $\phi=[\pi^{d/2}/2^{d-1}d\Gamma(d/2)] n\sigma^d$ is the solid volume fraction. According to the expression \eqref{n1}, the spatial dependence of $p$ is through its dependence on the number density $n$ (or equivalently, the volume fraction $\phi$) and the granular temperature $T$. Thus, in dimensionless form, Eq.\ \eqref{2.1} yields
\begin{equation}
\label{2.2}
\beta \frac{\partial_z \ln n}{\partial_z \ln T}=-\left( p^*+g^* \right),
\end{equation}
where $p^*\equiv p/nT$, $\beta\equiv \partial_\phi (\phi p^*)$, and $g^*\equiv \rho g/n\partial_zT<0$ is a dimensionless parameter measuring the gravity relative to the thermal gradient. In addition, since the mean flow velocity of the gas vanishes, the mass balance equation for the number density $n_0$  \cite{G09} implies $j_{0,z}=0$, where $j_{0,z}$ is the mass flux of intruders.

To close the determination of the thermal diffusion factor $\Lambda$, a constitutive equation for the mass flux $j_{0,z}$ is needed. To first order in the spatial gradients (Navier-Stokes approximation), the constitutive equation for the mass flux $j_{0,z}$ is \cite{GDH07}
\begin{equation}
\label{2} j_{0,z}=-\frac{m_0^2}{\rho}D_{0}
\partial_z n_0-\frac{m_0 m}{\rho}D\partial_z n-
\frac{\rho}{T}D^T\partial_zT,
\end{equation}
where $D_{0}$ is the kinetic diffusion coefficient, $D$ is the mutual diffusion coefficient, and $D^T$ is the thermal diffusion coefficient. The condition $j_{0,z}=0$ leads to the relation
\begin{equation}
\label{2.3}
D_0^*\frac{\partial_z \ln n_0}{\partial_z \ln T}+D^*\frac{\partial_z \ln n}{\partial_z \ln T}=-D^{T*},
\end{equation}
where we have introduced the reduced transport coefficients $D^{T*}\equiv (\rho \nu/n_0T)D^T$, $D_0^*\equiv (m_0^2\nu/\rho T)D_0$, and  $D^*\equiv (m_0\nu/n_0 T)D$. Here, $\nu=n\sigma^{d-1}\sqrt{2T/m}$ is an effective collision frequency.

The explicit form of $\Lambda$ can be finally obtained from Eqs.\ \eqref{2.2} and \eqref{2.3}. The result is    \cite{G08}
\begin{equation}
\label{4} \Lambda=\frac{\beta D^{T*}-(p^*+g^*)(D_{0}^*+ D^*)}{\beta D_{0}^*}.
\end{equation}
It is quite apparent that in order to assess the impact of the parameters of the system (masses, sizes and coefficients of restitution) and the volume fraction $\phi$ on the thermal diffusion $\Lambda$, the explicit forms of the diffusion coefficients $D^*$, $D_0^*$ and $D^{T*}$ are needed. This can be achieved by solving the Enskog kinetic equation by means of the Chapman-Enskog method \cite{CC70}.

\section{Enskog kinetic theory. First and second Sonine approximations to the diffusion transport coefficients}
\label{sec3}

In the tracer limit ($n_0/n\to 0$), it is expected that the state of the granular gas (the solvent) is not affected by the presence of tracer particles and that the mutual interactions of the latter can be neglected as compared with their collisions with the particles of the solvent. Consequently, at a kinetic theory level, the tracer limit implies that the velocity distribution function $f(\mathbf{r}, \mathbf{v}, t)$ of the granular gas obeys the (closed) nonlinear Enskog equation while the velocity distribution function $f_0(\mathbf{r}, \mathbf{v}, t)$ of the intruder obeys the Enskog-Lorentz equation \cite{GF09}.

Moreover, as in Ref.\ \cite{G08}, in order to maintain the granular medium in a fluidized state, an external energy source is coupled to each particle in the form of a thermal bath. Here, we consider the situation of energy supply through random kicks \cite{WK96}: the particles of the system are
submitted between collisions to an uncorrelated white noise. This external force is written as
${\boldsymbol {\cal F}}_i=m_i {\boldsymbol \xi}$, where the corresponding stochastic acceleration ${\boldsymbol \xi}$ is chosen to be the same for the intruder and the gas particles \cite{BT02}. The associated forcing term in the Enskog equation is represented by a Fokker-Planck collision operator \cite{NE98} of the form $-\frac{1}{2}(\zeta T/m)\partial^2/\partial v^2$, where $\zeta$ is the cooling rate associated with the granular temperature $T$. The generalization of the force to the \emph{inhomogeneous} case is essentially a matter of choice and here, for simplicity, we have assumed that the stochastic force has the same form as in the homogeneous case except that now $\zeta$ and $T$ are in general functions of space and time. This simple generalization has been widely used for ordinary gases in shearing problems \cite{GS03}. It must be emphasized that this kind of forcing, which has been shown to be relevant for some two-dimensional experimental configurations with a rough piston \cite{urbach}, has been usually employed in computer simulations to analyze different problems in the case of monodisperse systems \cite{thermostat}.

The application of the Chapman-Enskog method leads to the constitutive equation \eqref{2} for the mass flux where the transport coefficients $D$, $D_0$ and $D^T$ are given by
\begin{equation}
D^{T}=-\frac{m_0}{\rho d}\int \dd\mathbf{v}\mathbf{v}\cdot \boldsymbol{\mathcal{A}}_{0}\left(
\mathbf{v}\right) , \label{DT}
\end{equation}
\begin{equation}
D_{0}=-\frac{\rho}{m_{0}n_{0}d}\int \dd\mathbf{v}\mathbf{v}\cdot \boldsymbol{\mathcal{B}}_{0}\left(
\mathbf{v}\right) ,\label{D0}
\end{equation}
\begin{equation}
D=-\frac{1}{d}\int \dd\mathbf{v}\mathbf{v}\cdot \boldsymbol{\mathcal{C}}_{0}\left( \mathbf{v}\right).
\label{D}
\end{equation}
As for elastic collisions, the quantities $\boldsymbol{\mathcal{A}}_{0}\left(
\mathbf{v}\right)$, $\boldsymbol{\mathcal{B}}_{0}\left(\mathbf{v}\right)$ and
$\boldsymbol{\mathcal{C}}_{0}\left(\mathbf{v}\right)$ are the solutions of a set of coupled linear integral equations [see Eqs.\ (A13)-(A15) of Ref.\ \cite{G09} for the driven case].
The standard method consists of approximating the above quantities by Maxwellians (at different temperatures) times truncated Sonine polynomial expansions. For simplicity, usually only the lowest Sonine polynomial (first Sonine approximation) is retained \cite{G08,G09,GHD07} and the results obtained from this simple approach agrees in general well with Monte Carlo simulations \cite{GM03}. Exceptions to this agreement are extreme mass and size ratios in the range of large inelasticity, although these discrepancies could be mitigated in part if one considers higher-order terms in the Sonine polynomial expansion. This has been recently shown \cite{GF09} in the undriven case (without thermostat) at the level of the transport coefficient $D_0^*$ where the second Sonine approximation yields a dramatic improvement (up to 50 \%) over the first Sonine approximation for some extreme cases. For this reason, we evaluate here the complete set of diffusion coefficients $D_{0}^*$, $D^*$, and $D^{T*}$ up to the second Sonine approximation.
The procedure to determine the transport coefficients $D_{0}$, $D$ and $D^T$ follows similar mathematical steps as those made in the undriven case and so we refer the interested reader to Ref.\ \cite{GF09} for more specific details. Here, only the final results are displayed.

The first Sonine approximations $D_{0}^*[1]$, $D^{T*}[1]$, and $D^*[1]$ for the (reduced) coefficients $D_0^*$, $D^{T*}$ and $D^*$, respectively, are given by
\begin{equation}
\label{n2}
D_0^*[1]=\frac{\gamma}{\nu_D^*},
\end{equation}
\begin{equation}
\label{n3}
D^{T*}[1]=\frac{\gamma-Mp^*}{\nu_D^*}+\frac{(1+\omega)^d}{2\nu_D^*}\frac{M}{1+M}\chi_0
(1+\alpha_0),
\end{equation}
\begin{equation}
\label{5} D^{*}[1]=\frac{\phi}{\nu_D^*}
\frac{\partial \gamma}{\partial \phi}-\frac{M}{\nu_D^*}\beta
+\frac{1}{2\nu_D^*}
\frac{\gamma+M}{1+M}\frac{\phi}{T}\left(\frac{\partial\mu_0}{\partial
\phi}\right)_{T,n_0}(1+\alpha_{0}),
\end{equation}
where $\gamma\equiv T_0/T$ is the temperature ratio, $M\equiv m_0/m$ is the mass ratio, $\omega\equiv \sigma_0/\sigma$ is the size ratio, $\chi_0$ is the intruder-gas pair correlation function, $\mu_0$ is the chemical potential of the intruder and
\begin{equation}
\label{n4} \nu_D^*=\frac{2\pi ^{(d-1)/2}}{d\Gamma \left( \frac{d}{2}\right)}
\left(\frac{1+\omega}{2}\right)^{d-1}\frac{\chi _{0}}{1+M}\left( \frac{M+\gamma}{M}
\right)^{1/2}(1+\alpha _{0}).
\end{equation}
Since granular fluids lack a thermodynamic description, the concept of chemical potential appearing in Eq.\ \eqref{5} could be questionable. The presence of $\mu_0$ in our theory is justified because we want to recover the results derived from the revised Enskog kinetic equation for elastic mixtures \cite{MCK83} (see Appendix C of Ref.\ \cite{GHD07} for an explanation of the choice of some functional derivatives appearing in the inelastic Enskog theory). Given that the explicit form of the chemical potential must be known to evaluate the diffusion transport coefficients, for practical purposes, the expression considered here for $\mu_0$ is the same as the one obtained for ordinary mixtures ($\alpha=\alpha_0=1$). Although this derivation requires the use of thermodynamic relations that only apply for elastic systems, we expect that this approximation could be reliable for not too strong values of dissipation. More comparisons with computer simulations are needed to support the above expectation.

If we require the stochastic acceleration to be the same for both species, it is straightforward to show that he temperature ratio $\gamma$ fulfills \cite{BT02,DHGD02}
\begin{equation}
\label{n4.1}
\gamma \zeta_0^*=M\zeta^*,
\end{equation}
where in the Gaussian approximation the cooling rates $\zeta^*$ and $\zeta_0^*$ are
\begin{equation}
\label{n5}
\zeta^*=\frac{\sqrt{2}\pi ^{(d-1)/2}}{d\Gamma \left( \frac{d}{2}\right)} \chi (1-\alpha^2),
\end{equation}
\begin{eqnarray}
\label{n6}
\zeta_0^*&=&\frac{4\pi ^{(d-1)/2}}{d\Gamma \left( \frac{d}{2}\right)}
\left(\frac{1+\omega}{2}\right)^{d-1}\frac{\chi _{0}}{1+M}\left( \frac{M+\gamma}{M}
\right)^{1/2}\nonumber\\
&\times& (1+\alpha _{0})\left[1-\frac{M+\gamma}{2\gamma(1+M)}(1+\alpha_0)\right].
\end{eqnarray}
The derivative $\partial_\phi \gamma$ appearing in Eq.\ \eqref{5} for $D^*$ can be obtained by taking the derivative with respect to $\phi$ under the condition \eqref{n4.1}. This yields the relation
\begin{equation}
\label{n7}
\frac{\partial \gamma}{\partial \phi}=\frac{M \left(\frac{\partial \zeta^*}{\partial \chi}\right)
\left(\frac{\partial \chi}{\partial \phi}\right)-\gamma \left(\frac{\partial \zeta_0^*}{\partial \chi_0}\right)\left(\frac{\partial \chi_0}{\partial \phi}\right)}{\zeta_0^*+
\gamma \frac{\partial \zeta_0^*}{\partial \gamma}}.
\end{equation}

It must be also remarked that the steady state condition \eqref{n4.1} has been also obtained by considering local boundary conditions to the Enskog equation (see the Appendix of Ref.\ \cite{DHGD02}). More specifically, the boundary condition considered is a sawtooth vibration of one wall such that every particle encountering the wall has a reflected speed increased by twice the velocity of the wall in the component normal to the wall. In the limit that the wall velocity is large compared to the thermal velocities of each species (gas particles and intruder), the condition \eqref{n4.1} is recovered. This equivalence suggests that the results derived from the relation \eqref{n4.1} can be taken as plausible first approximations for qualitative comparisons with experimental results \cite{BT02}.

\begin{figure}
\includegraphics[width=0.7 \columnwidth,angle=0]{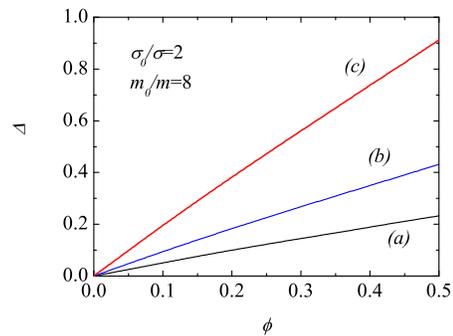}
\caption{(color online) Plot of $\Delta\equiv -\phi \partial_\gamma \phi$ versus the solid volume fraction $\phi$ for a hard-sphere gas in the case $m_0/m=8$ and $\sigma_0/\sigma=2$. Three different cases are considered: (a) $\alpha=\alpha_0=0.9$, (b) $\alpha=\alpha_0=0.8$, and (c) $\alpha=\alpha_0=0.5$.
\label{fign1}}
\end{figure}

The expression \eqref{5} for the diffusion coefficient $D^{*}[1]$ differs from the one derived previously in Ref.\ \cite{G08} by the presence of the term $\partial_\phi \gamma$. This contribution was implicitly neglected in those calculations. Equation \eqref{5} corrects this previous approximation.
In order to assess the effect of this new contribution, Fig.\ \ref{fign1} shows the dependence of $\Delta\equiv -\phi\partial_\phi \gamma$ on the solid volume fraction $\phi$ for three different values of the (common) coefficient of restitution $\alpha=\alpha_0$. We observe that, at a given value of $\alpha$, $\Delta$ increases with increasing $\phi$. In addition, at a given value of $\phi$, it is quite apparent that $\Delta$ also increases with increasing collisional dissipation. Therefore, the influence of the density dependence of the temperature ratio on thermal diffusion is expected to be important for moderate densities and strong dissipation.

\begin{figure}
\includegraphics[width=0.7 \columnwidth,angle=0]{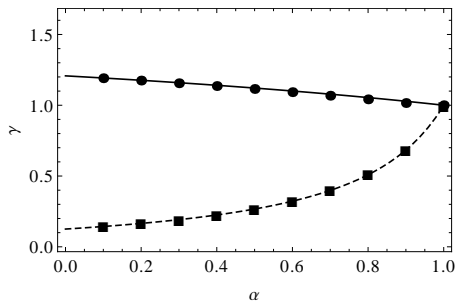}
\caption{Temperature ratio $\gamma\equiv T_0/T$ versus the (common) coefficient of restitution $\alpha_0=\alpha$ for a hard-sphere system ($d=3$) in the case $\sigma_0/\sigma=2$, $\phi=0.2$ and two values of the mass ratio $m_0/m$: $m_0/m=2$ (solid line and circles) and $m_0/m=\frac{1}{8}$ (dashed line and squares). The lines are the theoretical results and the symbols refer to the numerical results obtained from the DSMC method.   \label{fign2}}
\end{figure}
The evaluation of the second Sonine approximations $D_{0}^*[2]$, $D^{T*}[2]$, and $D^*[2]$ is more involved. As mentioned before, previous calculations carried out in the undriven case for the second Sonine approach \cite{GF09} allows one to easily extend these expressions when the gas is heated by means of an stochastic thermostat. Their explicit forms are provided in the Appendix \ref{appA}. In general, the forms of $D_{0}^*[2]$, $D^{T*}[2]$, and $D^*[2]$ have a complex dependence on the parameter space of the problem (the mass and size ratios, the solid fraction and the coefficients of restitution). In the elastic limit
($\alpha=\alpha_0=1$) of a three-dimensional gas, Eqs.\ \eqref{a1}--\eqref{a3} for the above coefficients
agree with those previously obtained for a gas mixture of elastic hard spheres \cite{M54}.
Moreover, in the case of mechanically equivalent particles ($m_0=m, \sigma_0=\sigma, \alpha_0=\alpha$),
as expected, one obtains $D^{T*}[2]=0$ and $D_{0}^*[2]=-D^*[2]$. Both limit cases
confirm the self-consistency of the expressions derived here.
\begin{figure*}
\begin{center}
\begin{tabular}{lr}
\resizebox{6.5cm}{!}{\includegraphics{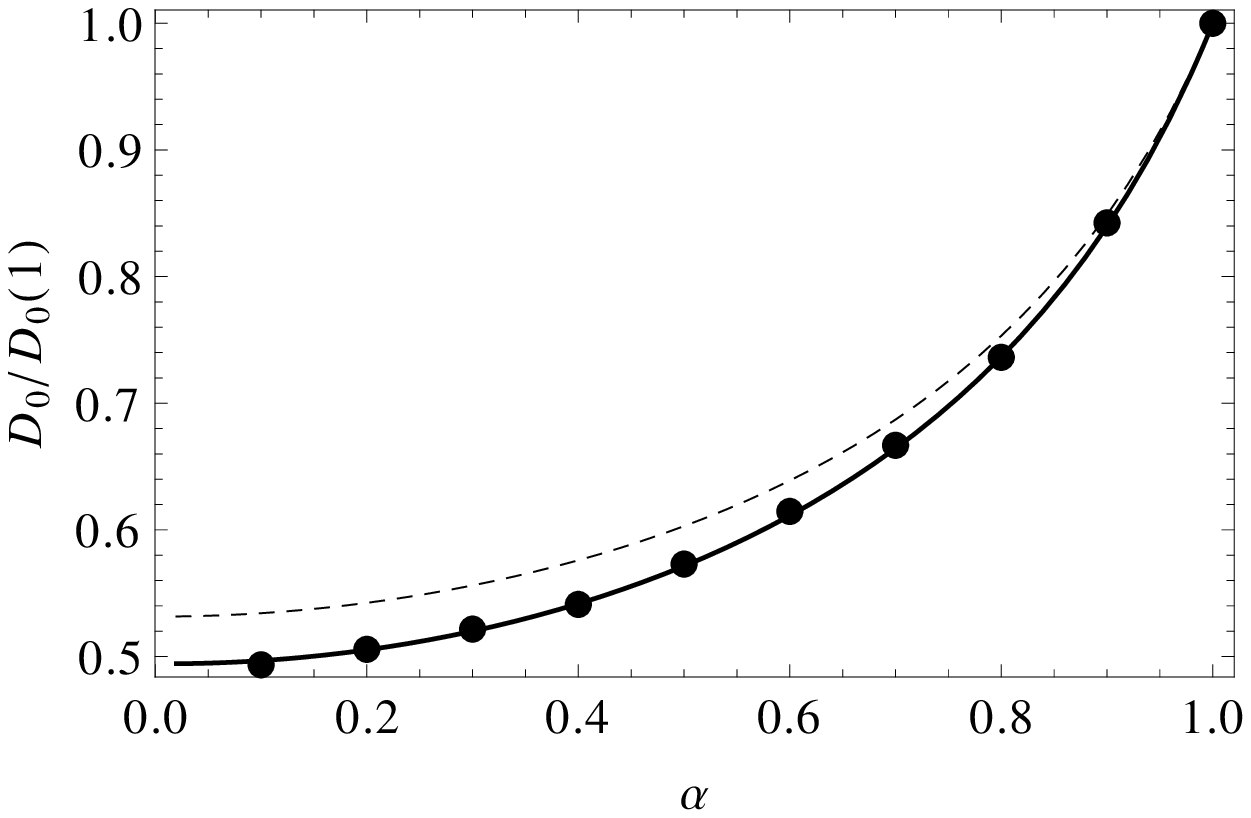}}&\resizebox{6.5cm}{!}
{\includegraphics{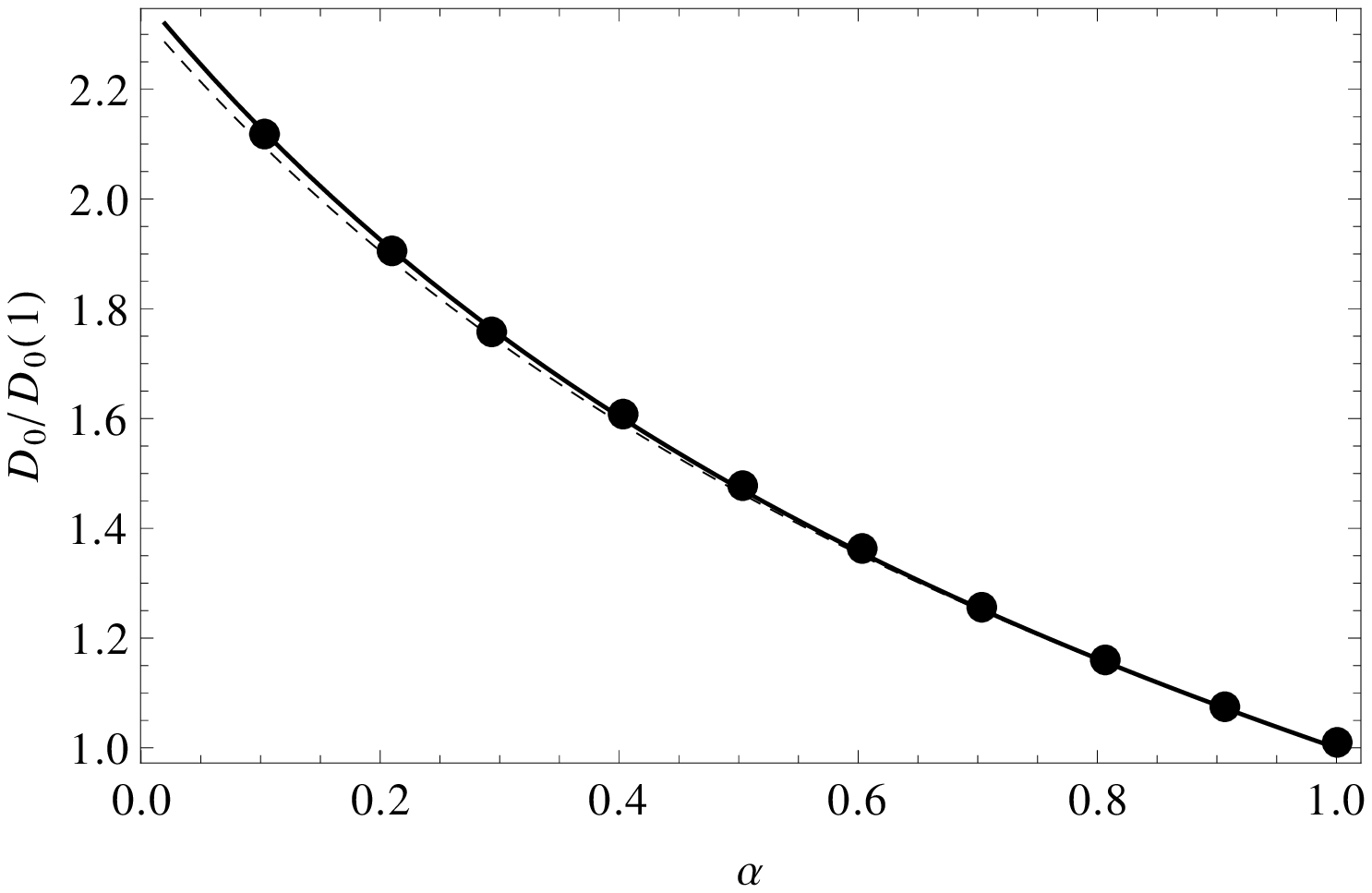}}
\end{tabular}
\end{center}
\caption{Reduced kinetic diffusion coefficient $D_0(\alpha)/D_0(1)$ as a function of the (common) coefficient of restitution $\alpha=\alpha_0$ for a system of hard spheres with $\omega=2$ and $\phi=0.2$. The left panel is for $M=1/8$ while the right panel is for $M=2$. The solid lines correspond to the second Sonine approximation and the dashed lines refer to the first Sonine approximation. The symbols are the results obtained from Monte Carlo simulations. Here, $D_0(1)$ is the elastic value of the kinetic diffusion coefficient consistently obtained in each approximation.
\label{mu2}}
\end{figure*}
\begin{figure*}
\begin{center}
\begin{tabular}{lr}
\resizebox{6.5cm}{!}{\includegraphics{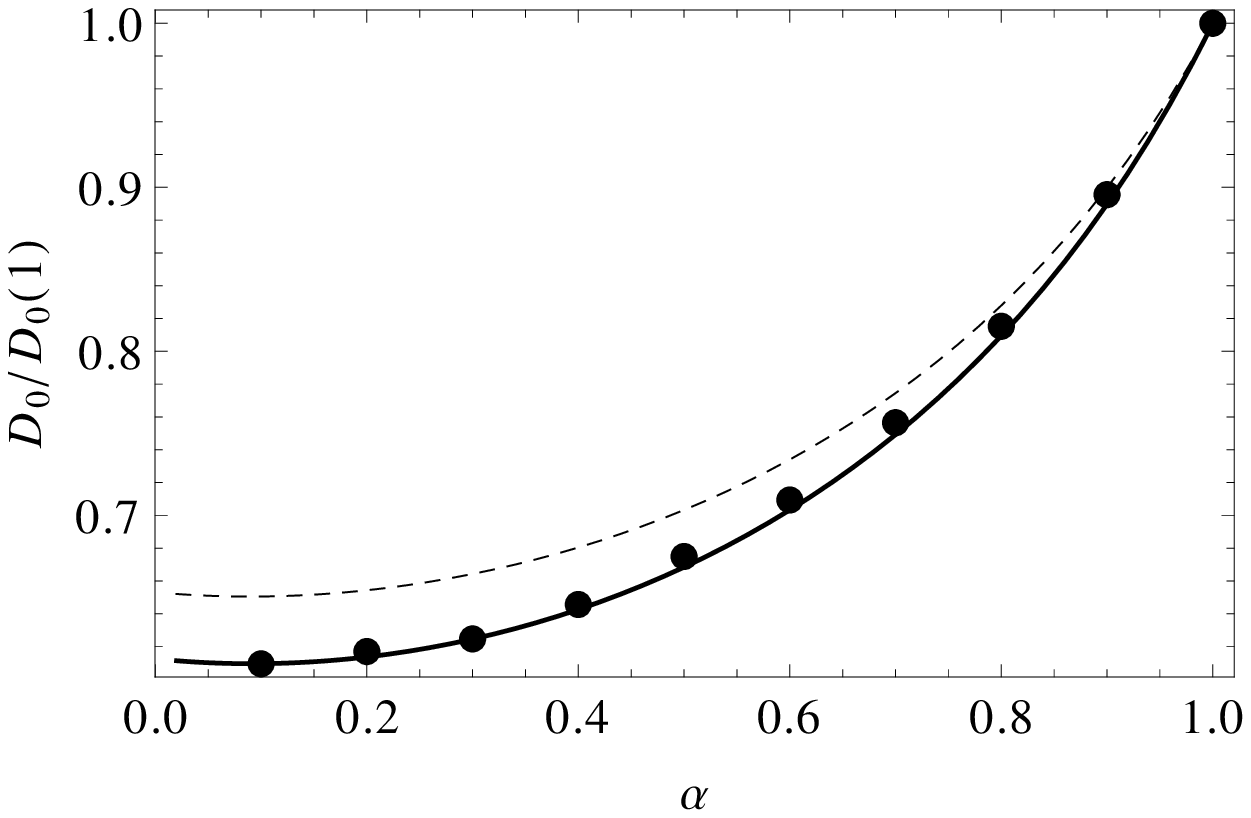}}&\resizebox{6.5cm}{!}
{\includegraphics{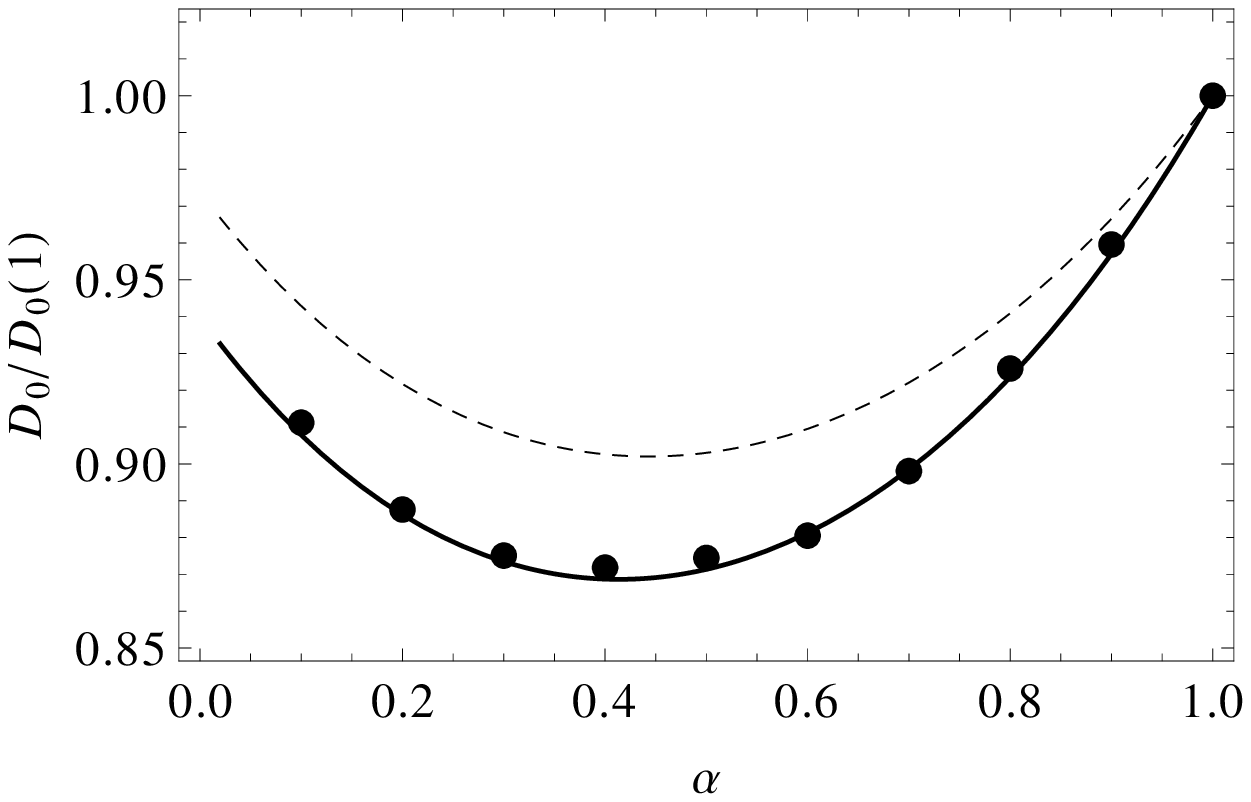}}
\end{tabular}
\end{center}
\caption{Reduced kinetic diffusion coefficient $D_0(\alpha)/D_0(1)$ as a function of the (common) coefficient of restitution $\alpha=\alpha_0$ for a system of hard spheres with $\omega=2$ and $\phi=0.2$. The left panel is for $M=1/5$ while the right panel is for $M=1/2$. The solid lines correspond to the second Sonine approximation and the dashed lines refer to the first Sonine approximation. The symbols are the results obtained from Monte Carlo simulations. Here, $D_0(1)$ is the elastic value of the kinetic diffusion coefficient consistently obtained in each approximation.
\label{mu15}}
\end{figure*}

\section{Comparison with Monte Carlo simulations}
\label{sec4}

It is important to note that, unless the Sonine expansion is convergent, the introduction of the second order in the Sonine polynomial expansion does not guarantee \emph{a priori} the improvement of the analytical results. In order to gauge the theoretical predictions one would have to compare the latter with computer simulations. This is the main goal of this Section where the expressions \eqref{n2} and \eqref{a1} of the first and second Sonine approximations, respectively, for the kinetic diffusion coefficient $D_0$ will be compared with Monte Carlo simulations.

The diffusion coefficient $D_0$ of intruders has been extracted by solving numerically the \emph{homogeneous} Enskog equation by means of the DSMC method. As in the undriven case \cite{GF09,GM04}, the coefficient $D_0$ can be obtained from the mean square displacement of the intruder after a time interval $t$ as \cite{L89}
\begin{equation}
\label{3.1}
\frac{\partial}{\partial t}\langle |{\bf r}(t)-{\bf r}(0)|^2 \rangle =\frac{2dD_0}{n},
\end{equation}
where $|{\bf r}(t)-{\bf r}(0)|$ is the distance traveled by the intruder from $t=0$
until time $t$. Equation (\ref{3.1}) is the Einstein form of the diffusion
coefficient. This relation can be used also in Monte Carlo simulations of granular
gases to measure the diffusion coefficient. In an
unbounded system like ours, the DSMC method has two steps that are repeated in each
time iteration. In the first step (free streaming stage), the velocity ${\bf v}_i$ of every particle (intruder and gas particles) is changed to ${\bf v}_i+{\bf w}_i$, where ${\bf w}_i$ is due to the stochastic force $\boldsymbol{\mathcal{F}}_i$, and is randomly drawn from a Gaussian probability distribution (see Ref.\ \cite{MS00} for more details). The second step accounts for the collisions among particles. Because the tracer limit ($n_0/n\to 0$), during our simulations, collisions between intruder particles themselves are not considered, and when a collision between the intruder and a particle of the gas takes place, the post-collisional velocity obtained from the scattering rule is only assigned to the intruder. According to this scheme, the numbers of particles have simply a statistical meaning, and hence they can be chosen arbitrarily.

The extension of the DSMC method to study
the diffusion of intruders in a \emph{dense} homogeneous granular gas requires the changes $J[f,f]\to \chi J[f,f]$ and $J_0[f_0,f]\to \chi_0 J_0[f_0,f]$. Here, $J[f,f]$ and $J_0[f_0,f]$ refer to the (closed) Boltzmann and Boltzmann-Lorentz collision operators, respectively \cite{VSG11}. For the DSMC method to work appropriately, the time step needs to be small in comparison with the microscopic
time scale of the problem (which is set by the inverse of the collision frequency $\nu$) and we also
need a sufficiently high number of simulated particles \cite{B94}. We have used
in the simulations of this paper a time step $\delta t=2.5\times10^{-4}\nu^{-1}$ and
$N=2\times 10^6$ simulated particles for each species.

Before studying the diffusion coefficient $D_0$, it is noteworthy to compare the theoretical predictions for the temperature ratio $\gamma$ with computer simulations. Figure \ref{fign2} shows $\gamma$ versus the (common) coefficient of restitution $\alpha_0=\alpha$ for a three-dimensional gas for $\omega=2$, $\phi=0.2$ and two different values of the mass ratio: $M=2$ and $M=1/8$. For hard spheres, a good approximation for the pair correlation function $\chi$ is provided by the Carnahan-Starling form \cite{CS69}
\begin{equation}
\label{7.2} \chi=\frac{1-\frac{1}{2}\phi}{(1-\phi)^3},
\end{equation}
while the intruder-gas pair correlation function is given by \cite{B70}
\begin{equation}
\label{7.3}
\chi_0=\frac{1}{1-\phi}+3\frac{\omega}{1+\omega}\frac{\phi}{(1-\phi)^2}+2
\frac{\omega^2}{(1+\omega)^2}\frac{\phi^2}{(1-\phi)^3}.
\end{equation}
It is quite apparent the excellent agreement found between the theory [obtained from the condition \eqref{n4.1}] and simulation, showing again  the accuracy of the approximations \eqref{n5} and \eqref{n6} to estimate the temperature ratio. As expected \cite{BT02}, the heavier particles carry generically more kinetic energy than the lighter ones. Moreover, the deviations from the energy equipartition increase as the mass differences between intruders and particles of the gas increase. As we will show later, in general the effect of the temperature differences on the thermal diffusion segregation is quite significant.

Let us consider now the kinetic diffusion coefficient. Figures \ref{mu2} and \ref{mu15} show the ratio $D_0(\alpha)/D_0(1)$ as a function of the (common) coefficient of restitution $\alpha=\alpha_0$ for $\omega=2$ and $\phi=0.2$ in the case of spheres ($d=3$). We have reduced $D_0(\alpha)$ with respect to its elastic value $D_0(1)$ consistently obtained in each Sonine approximation. The solid lines are the theoretical results derived from the second Sonine approximation while the dashed lines refer to the first Sonine approximation. We observe that in general, while the second Sonine approximation agrees very well with simulation data, some disagreement appears with the first Sonine approximation for strong dissipation when the intruder is lighter than gas particles. In this case, the first Sonine approximation underestimates the kinetic diffusion coefficient. On the other hand, in the opposite case (when the intruder is heavier than gas particles) the first and second Sonine approximations are practically indistinguishable in the complete range of values of $\alpha$ explored and both approaches provide a good agreement with Monte Carlo simulations. All these results clearly confirm the accuracy of the second Sonine approximation for the coefficient $D_0$, even for low values of the coefficient of restitution. Similar conclusions were obtained in the undriven case \cite{GF09} for the diffusion coefficient, although the quantitative differences between both Sonine solutions were more significant than the ones observed here.

\section{Phase diagrams for hard spheres}
\label{sec5}

The explicit dependence of the thermal diffusion factor $\Lambda$ on the parameter space of the problem can be obtained when one substitutes the expressions of the transport coefficients $D_{0}^*$, $D^*$, and $D^{T*}$ [Eqs.\ \eqref{n2}--\eqref{5} for the first Sonine approximation and Eqs.\ \eqref{a1}--\eqref{a3} for the second Sonine approximation] and the pressure $p^*$ (and its corresponding derivative $\beta$) into Eq.\ \eqref{4}. It is quite evident that the influence of the parameters of the mixture (masses, diameters, density and coefficients of restitution) and the (reduced) gravity on the sign of $\Lambda$ is rather complicated, given the large number of parameters involved. In particular, the condition $\Lambda=0$ provides the criterion for the BNE/RBNE transition. Given that the results show that both $\beta$ and $D_0^*$ are positive, then, according to Eq.\ \eqref{4}, the transition criterion is
\begin{equation}
\label{7}
\beta D^{T*}=(p^*+g^*)(D_{0}^*+D^*).
\end{equation}
Note that the thermal diffusion factor is in general a \emph{non-uniform} function since it depends on $z$ through its dependence on the volume fraction $\phi(z)$ and the thermal gradient $\partial_zT$ (in the case that the temperature profile is \emph{not} linear). However, since the expression \eqref{4} for $\Lambda$ has been obtained up to the Navier-Stokes order (first order in the spatial gradients), our segregation criterion only \emph{strictly} applies for regions where the density and thermal gradients are quite small. Under these conditions, one can assume that the thermal diffusion factor is practically constant, and so the criterion \eqref{7} can be considered as a global feature of the system. In the case of \emph{arbitrary} spatial gradients, the numerical solution (beyond the Navier-Stokes domain) of the Enskog equation via the DSMC method would give better quantitative agreement with molecular dynamics simulations or experiments than the Navier-Stokes results reported here.
\begin{figure}
\includegraphics[width=0.7 \columnwidth,angle=0]{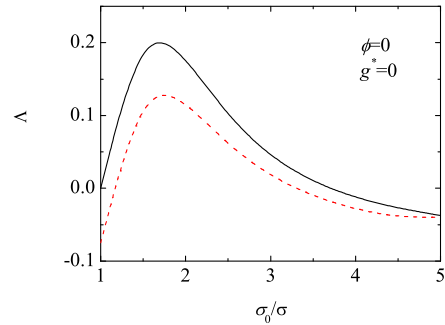}
\caption{(color online) Plot of the thermal diffusion factor $\Lambda$ obtained from the second Sonine approximation as a function of the diameter ratio $\sigma_0/\sigma$ for a dilute ($\phi=0$) hard-sphere gas in the absence of gravity ($g^*=0$) when the intruder and gas particles have the same mass density ($m_0/m=(\sigma_0/\sigma)^3$). Two different cases are considered: $\alpha=\alpha_0=0.7$ (solid line) and $\alpha=0.5$, $\alpha_0=0.9$ (dashed line).   \label{fig2}}
\end{figure}
\begin{figure}
\includegraphics[width=0.7 \columnwidth,angle=0]{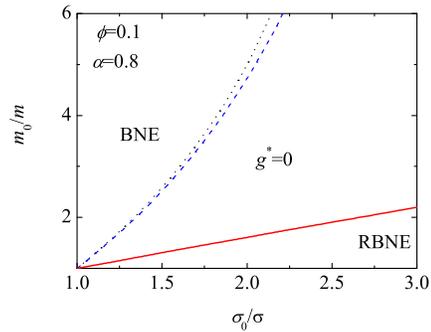}
\caption{(color online) Phase diagram for BNE/RBNE at $\phi=0.1$ in the absence of
gravity with $\alpha=0.8$. Points
above the curve correspond to $\Lambda>0$ (BNE) while points below the curve correspond
to $\Lambda<0$ (RBNE). The dotted line is the result obtained in Ref.\ \cite{G08}, the dashed line is the result obtained here from the first Sonine approximation while the solid line is the result derived from the second Sonine approximation.  \label{fig3}}
\end{figure}
\begin{figure}
\includegraphics[width=0.7 \columnwidth,angle=0]{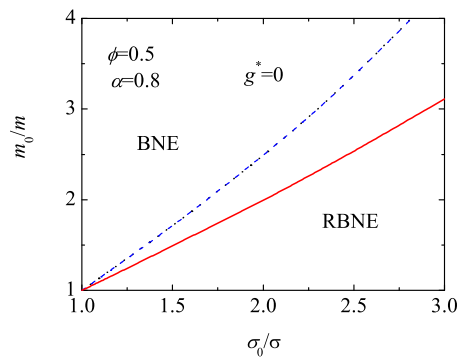}
\caption{(color online) The same as in Fig.\ \ref{fig3} but at $\phi=0.5$. \label{fig4}}
\end{figure}

As expected, if the impurities are mechanically equivalent to the host gas, the system is monodisperse and thermal segregation does not occur ($\Lambda=0$). This is consistent with our results since in this limit case $D^{T*}[2]=0$ and $D_{0}^*[2]=-D^*[2]$, so that the condition \eqref{7} applies for any value of $\phi$ and $\alpha$. For a dilute gas ($\phi=0$), the first Sonine approximation to Eq.\ \eqref{7} simply yields
\begin{equation}
\label{7.1}
g^*(\gamma-M)=0,
\end{equation}
and so no segregation occurs in the absence of gravity. This result is consistent with the criterion obtained from the Boltzmann equation \cite{G06}. However, when the second Sonine correction to the diffusion coefficients is retained, the results show that thermal segregation appears ($\Lambda\neq 0$) for a dilute gas even in the absence of gravity. This is illustrated in Fig.\ \ref{fig2} showing a non monotonic dependence of $\Lambda$ on the diameter ratio. In addition, when $g^*\neq 0$, Eq.\ \eqref{7.1} leads to the criterion $\gamma=M$. This condition also agrees with the Boltzmann results \cite{BRM05} of the undriven case.

To assess the impact of the Sonine approximation on segregation, henceforth we only consider
the physical case of hard spheres ($d=3$) with a common coefficient of restitution ($\alpha=\alpha_0$).
This reduces the parameter space of the problem to five dimensionless parameters:
$g^*$, $\sigma_0/\sigma$, $m_0/m$, $\phi$ and $\alpha$. In the case of hard spheres,
The expression for the chemical potential of the intruder consistent with the approximation (\ref{7.3}) is
\cite{RG73}
\begin{eqnarray}
\label{7.4} \frac{\mu_0}{T}&=&\ln (n_0\lambda_0^3)-\ln (1-\phi)+3\omega \frac{\phi}{1-\phi}\nonumber\\
& & +
3\omega^2\left[\ln (1-\phi)+\frac{\phi(2-\phi)}{(1-\phi)^2}\right]\nonumber\\
& & -\omega^3\left[2\ln (1-\phi)+\frac{\phi(1-6\phi+3\phi^2)}{(1-\phi)^3}\right],
\end{eqnarray}
where $\lambda_0(T)$ is the (constant) de Broglie's thermal wavelength \cite{RG73}. The forms \eqref{7.2}, \eqref{7.3}, and \eqref{7.4} for $\chi$, $\chi_0$, and $\mu_0$, respectively, are the same as those used in our previous calculations \cite{G08,G09}.

Figure \ref{fig3} shows a phase diagram in the $(m_0/m, \sigma_0/\sigma)$ plane for a low-density gas ($\phi=0.1$) in the absence of gravity ($g^*=0$) for $\alpha=0.8$. This situation (thermal gradient dominates over gravity) can be achieved in experiments of granular mixtures subjected to horizontal vibration where the role of gravity can be ignored \cite{exp}. The corresponding phase diagram obtained from the first Sonine approximation by assuming $\partial_\phi \gamma=0$ \cite{G08} is also plotted for comparison. We observe first that the latter assumption has no significant effect on the first Sonine prediction since both results (dotted and dashed lines) practically coincide in the range of diameter ratios studied. On the other hand, although the first Sonine approximation reproduces qualitatively well the trends of the phase diagram, the former overestimates dramatically the predictions of the second Sonine approximation, especially at large mass and size ratios. The results also show that the quantitative discrepancies between both Sonine solutions decrease as the volume fraction $\phi$ increases. This is illustrated in Fig.\ \ref{fig4} for the same system as in Fig.\ \ref{fig3}. The above results suggest that the Sonine expansion for thermal diffusion exhibits a poor convergence when gravity is absent and the gas is moderately dense (say for instance, $\phi \lesssim 0.2$). In particular, the first Sonine approximation turns out to be a very poor approximation to $\Lambda$ in the case of dilute gases. This conclusion agrees with the results obtained many years ago by Kincaid \emph{et al.} \cite{KCM87} for ordinary binary mixtures since they concluded that the second Sonine approximation is much better approximation than the first one. On the other hand, given that the first Sonine approximation to $D_0$ agrees well with simulation data (see the right panel of Fig.\ \ref{mu2}) in this parameter region (i.e., for mass ratios larger than one), the poor convergence of the Sonine approximation in this region is mainly due to the coefficients $D$ and $D^T$.

\begin{figure}
\includegraphics[width=0.7 \columnwidth,angle=0]{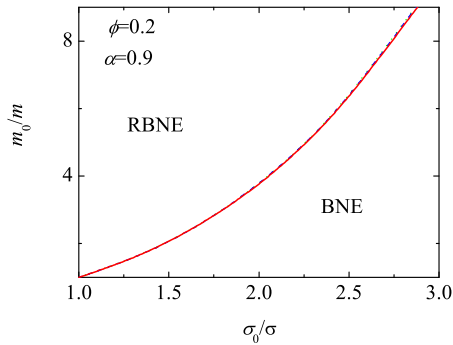}
\caption{(color online) Phase diagram for BNE/RBNE for $\phi=0.2$ in the absence of
a thermal gradient ($|g^*|\to \infty$) with $\alpha=0.9$. The dotted and dashed lines refer to the results obtained in Ref.\ \cite{G08} and here, respectively, from the first Sonine approximation while the solid line is the result derived from the second Sonine approximation.  \label{fig5}}
\end{figure}

We consider now the opposite limit $|g^*|\to \infty$, namely, when the temperature of the bed is assumed uniform so that the segregation of intruder is essentially driven by gravity. This situation (gravity dominates over the temperature gradient) has been previously studied by several authors \cite{JY02,TAH03} by using kinetic theory and by means of computer simulations \cite{HQL01} and experiments \cite{BEKR03}.
In this limit, the condition \eqref{7} reduces to $D_0^*+D^*=0$. Figure \ref{fig5} shows a phase diagram for $\phi=0.2$ and $\alpha=0.9$ for this limit case. It is apparent that, in contrast to the case $g^*=0$, the RBNE regime appears essentially now for both large mass ratio and/or small diameter ratio. Regarding the influence of the assumption $\partial_\phi \gamma=0$ in the first Sonine approximation \cite{G08} and of the order of the Sonine approximation used, Fig.\ \ref{fig4} shows clearly that the form of the phase diagram is practically independent of the approach used since the three curves collapse in a common curve.

Next, we analyze the influence of inelasticity on thermal diffusion by considering only the most accurate approach (the second Sonine approximation). To illustrate this effect, two values of $\alpha$ are considered in Fig.\ \ref{fig6} for $\phi=0.1$ and $g^*=0$. We see that, in the absence of gravity, the main effect of collisional dissipation is to reduce the size of the BNE, in contrast to what happens for finite concentration \cite{G11}. In addition, comparison with the results obtained for $\alpha=0.5$ assuming energy equipartition ($T_0=T$) shows that the impact of nonequipartition of granular energy on segregation is important (when $g^*=0$), especially as the size ratio increases. However, the influence of nonequipartition is smaller than the one previously
found from the first Sonine approximation (see for instance, Fig.\ 2 of Ref.\ \cite{G08}). Moreover, a comparison of the second Sonine approximation (not shown here) with previous theories \cite{JY02,TAH03} in the limit $|g^*|\to \infty$ shows similar discrepancies as those previously reported (see Fig.\ 4 of Ref.\  \cite{G08}). This is consistent with the results presented in Fig.\ \ref{fig3},
where the first and second Sonine solutions are practically indistinguishable.
\begin{figure}
\includegraphics[width=0.7 \columnwidth,angle=0]{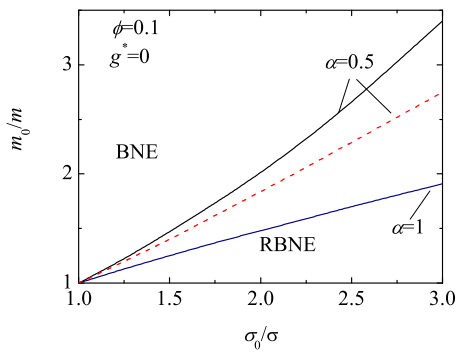}
\caption{(color online) Phase diagram for BNE/RBNE at $\phi=0.1$ with $g^*=0$ and two different values of $\alpha$. The solid lines are the results from the second Sonine approximation while the dashed line is the result derived from the latter approach for $\alpha=0.5$ but assuming $T_0=T$.  \label{fig6good}}
\end{figure}

\section{Summary and discussion}
\label{sec6}

In spite of its practical importance, the understanding of particle segregation within polydisperse, rapid granular flows is still an open problem. This is due essentially to the complexities that arise during the derivation of kinetic theory models. Most of the previous theoretical works are based on theories that consider systems constituted by nearly elastic particles \cite{JY02,TAH03,HH96,AW98}, assume an equipartition of granular energy \cite{JY02,HH96,AW98} and/or they are restricted to dilute gases \cite{BRM05,G06,SGNT06}. This paper has addressed the problem of segregation by thermal diffusion of an intruder in a driven granular gas. The analysis is based on Navier-Stokes constitutive equations with diffusion coefficients derived from the revised Enskog kinetic theory \cite{GDH07}.
The theory is not restricted to \emph{nearly} elastic spheres, considers the influence of the
nonequipartition of granular energy on segregation, and applies to \emph{moderate} values of the solid volume fraction.
In addition, in contrast to the previous attempt carried out by one of the authors of the present paper \cite{G08}, the analysis incorporates the density dependence of the temperature ratio ($\partial_\phi \gamma$) and considers the second Sonine approximation (two polynomials in the Sonine polynomial expansion) to the transport coefficients. These new results are the most significant contribution of the present work. In this context, this paper complements and extends previous papers on segregation of an intruder in driven \cite{G08,G09} and undriven \cite{GF09} dense granular gases.

In the steady state with gradients only along the vertical direction, the \emph{sign} of the thermal diffusion factor $\Lambda$ [defined by Eq.\ \eqref{1}] provides information on the tendency of the intruder to move towards the colder (BNE) or hotter (RBNE) plate. The factor $\Lambda$ has been evaluated by following two complementary approaches. First, by using the momentum balance equation along with the constitutive equation \eqref{2} for the mass flux of the intruder, $\Lambda$ is expressed in terms of the pressure $p$ (and its derivative with respect to the solid volume fraction $\phi$) and the transport coefficients $D_0$, $D^T$, and $D$. Then, the form of the diffusion transport coefficients has been determined by solving the Enskog equation from the Chapman-Enskog method up to the second Sonine approximation. This finally gives $\Lambda$ as a function of the mass and the diameter ratios, the volume fraction, the coefficients of restitution and the (reduced) gravity $g^*$ (a parameter measuring the gravity relative to the thermal gradient).

In order to check the reliability of the first and second Sonine approximations, a comparison with Monte Carlo simulations of the Enskog equation for the kinetic diffusion coefficient $D_0$ has been performed. As for undriven gases \cite{GF09}, the comparison with simulation data shows the superiority of the second Sonine solution over the first one specially when the gas particles are heavier than the intruder. However, the discrepancies found here between both Sonine approximations are less important than those previously reported for undriven systems \cite{GF09}.

The condition $\Lambda=0$ [see Eq.\ \eqref{7}] provides the segregation criterion for the transition BNE
$\Leftrightarrow$ RBNE. A systematic study of the form of the BNE/RBNE phase diagrams in the mass and
diameter ratio plane has been carried out in Sec.\ \ref{sec5} for hard spheres ($d=3$) in the case of a
(common) coefficient of restitution $\alpha=\alpha_0$. Regarding the form of the phase diagrams,
we observe first that the impact of the term $\partial_\phi \gamma$ on the first Sonine approximation
is very small (see, for instance, Figs.\ \ref{fig3} and \ref{fig4}) so that the conclusions made in
Ref.\ \cite{G08} are not practically altered by the presence of this new term. With respect to the
effect of the Sonine approximation considered, the results reported here show that in general the
influence of the order of the Sonine solution is much more significant in the absence of gravity
($g^*=0$) than in the opposite limit ($|g^*|\to \infty$), where both approximations yield practically
the same diagrams (see, for instance, Fig.\ \ref{fig3}). In particular, when $g^*=0$, the first
Sonine approach clearly overestimates the RBNE region (see Fig.\ \ref{fig3}), while it predicts a
bigger influence of collisional dissipation on the phase diagram than the one obtained from the
second Sonine correction (compare Fig.\ 2 of Ref.\ \cite{G08} with Fig.\ \ref{fig4} of the present paper).
Although the accuracy of the second Sonine approximation to the diffusion coefficients has been only
tested in the case of the coefficient $D_0$, it can be reasonably expected that the segregation criterion
derived here
in the second Sonine approximation for a heated gas compares better with computer simulations
than segregation criteria from previous works in the Navier-Stokes domain (small gradients
of density and temperature). A previous comparison \cite{BRM05} for dilute gases confirms this expectation. Given that the present results apply to moderate densities, it is hoped that this paper could stimulate the performance of such simulations.

An important question addressed partially in this paper is about the convergence of the Sonine polynomial expansions considered here. This is a quite difficult question, specially in the case of granular fluids where the studies of the impact of higher-order terms on transport are more scarce than for ordinary gases. In this latter case, for instance the analysis of transport properties of dense binary mixtures with one tracer component \cite{MC84} indicates that the convergence of the Sonine expansion improves significantly with increasing values of the mass ratio $M$. Similar trends have been found here at the level of the tracer diffusion coefficient $D_0$. Unfortunately, the lack of available simulation data for the remaining diffusion transport coefficients $D$ and $D^T$ in the driven or undriven cases prevent us to assess the reliability of the first and second Sonine approximations. Since those coefficients are also involved in the expression of the thermal diffusion factor $\Lambda$, no definitive conclusions on the accuracy of their second Sonine forms can be drawn. However, the results displayed in Section \ref{sec5} for thermal diffusion segregation seem to indicate that while the convergence of the Sonine expansion for $\Lambda$ is quite good when $|g^*|\to \infty$ (see for instance Fig.\ \ref{fig5} where both Sonine approaches lead practically to the same phase diagrams), it does not happen the same for a low-density gas in the opposite limit ($g^*=0$) since there is an abrupt change in the form of the phase diagram for BNE/RBNE (see for instance Fig.\ \ref{fig3}). In this situation one should perhaps consider higher-order polynomial terms (even beyond the second Sonine approximation) or one should consider alternative analytical routes, such as the so-called modified Sonine approximation \cite{GSM07}. This latter method is based on a modified version of the first Sonine approximation which replaces the Gaussian distribution weight function (used in the standard Sonine method) by the homogeneous cooling state distribution. In any case, more comparisons between segregation results derived in the driven case between computer simulations and the different approximate theories are needed before quantitative conclusions can be offered on the reliability of those kinetic theories.

\acknowledgments
The present work has been supported by the Ministerio de
Educaci\'on y Ciencia (Spain) through grants No. FIS2010-16587 (VG and FV) and No. MAT2009-14351-C02-02 (FV), partially financed by FEDER funds and by the Junta de Extremadura (Spain) through Grant No. GR10158.

\appendix
\section{Second Sonine expressions for the diffusion coefficients}
\label{appA}

The explicit expressions of the second Sonine approximations to the diffusion transport coefficients are displayed in this Appendix. They are given by
\begin{equation}
\label{a1}
D_{0}^*[2]=\frac{\nu_4^*\gamma}{\nu_1^*\nu_4^*-\nu_2^*(\nu_5^*-\zeta^*)},
\end{equation}
\vspace{0.2mm}
\begin{equation}
\label{a2}
D^{T*}[2]=\frac{\nu_4^*(X_1^*-a^*\gamma^2\nu_3^*)-\gamma^2\nu_2^*(X_2^*-a^*\nu_6^*)}
{\nu_1^*\nu_4^*-\nu_2^*(\nu_5^*-\zeta^*)},
\end{equation}
\begin{equation}
\label{a3}
D^{*}[2]=\frac{\nu_4^*(Y_1^*-c^*\gamma^2\nu_3^*)-\gamma^2\nu_2^*(Y_2^*-c^*\nu_6^*)}
{\nu_1^*\nu_4^*-\nu_2^*(\nu_5^*-\zeta^*)},
\end{equation}
where
\begin{equation}
\label{a4} X_1^*=-\left(M p^*-\gamma\right)+\frac{1}{2} (1+\omega)^d
{\cal M}_{0}\chi_0\phi (1+\alpha_0),
\end{equation}
\begin{eqnarray}
\label{a5} X_2^*&=&\gamma^{-1}+\frac{1}{2(d+2)} \frac{{\cal M}_{0}^2}{{\cal M}} (1+\omega)^d
\gamma^{-3}\chi_0\phi (1+\alpha_0)\nonumber\\
&\times &  \left\{\frac{{\cal M}
\gamma}{{\cal M}_0}\left[(d+2)({\cal M}_{0}^2-1)+(2d-5-9\alpha_0){\cal M}_{0}{\cal M}\right.\right.
\nonumber\\
&+ & \left.\left.(d-1+3\alpha_0+6\alpha_0^2){\cal M}^2\right]+6 {\cal M}^2(1+\alpha_0)^2\right\},\nonumber\\
\end{eqnarray}
\begin{equation}
\label{a6} Y_1^*=\phi \frac{\partial \gamma}{\partial \phi}-M\beta+
\frac{\gamma+M}{2(1+M)}\frac{\phi}{T}\left(\frac{\partial\mu_0}{\partial
\phi}\right)_{T,n_0}(1+\alpha_{0}),
\end{equation}
\begin{eqnarray}
\label{a7} Y_2^*&=&\frac{1}{2(d+2)}\frac{{\cal M}^2}{{\cal M}_{0}}\phi (1+\alpha_0)\frac{\partial}{\partial \phi}
\left(\frac{\mu_0}{T}\right)_{T,n_0} \nonumber\\
& & \times \left\{ \left[(d+8){\cal M}_{0}^2+(7+2d-9\alpha_0){\cal M}_{0}{\cal M}+(2+d
\right.\right. \nonumber\\
& &
\left.+3\alpha_0^2-
3\alpha_0){\cal M}^2\right]\theta+3{\cal M}^2(1+\alpha_0)^2\theta^3
\nonumber\\
& & +\left[(d+2){\cal M}_{0}^2+(2d-5-9\alpha_0){\cal M}_{0}{\cal M}+(d-1
\right.\nonumber\\
& & \left.\left.+3\alpha_0+6\alpha_0^2)
{\cal M}^2\right]\theta^2-(d+2)\theta(1+\theta)\right\}.
\end{eqnarray}
Here, ${\cal M}=m/(m+m_0)$, ${\cal M}_0=m_0/(m+m_0)$ and $\theta=m_0T/mT_0$ is the mean-square velocity of the gas particles relative to that of the intruder particle. In addition, when the gas is driven by the stochastic thermostat, the coefficients $a^*$ and $c^*$ are given by \cite{GM02}
\begin{equation}
\label{a8}
a^*=\nu_\kappa^{*-1}\left(1+3\frac{2^{d-3}}{(d+2)}\phi\chi (1+\alpha)^2(2\alpha-1)\right),
\end{equation}
\begin{equation}
\label{a9}
c^*=-3\nu_\kappa^{*-1}\frac{2^{d-3}}{(d+2)}\phi \left(2\chi+\phi\partial_\phi \chi\right)\alpha(1-\alpha^2),
\end{equation}
where
\begin{eqnarray}
\label{a10} \nu_\kappa^*&=&\frac{8}{d(d+2)}\frac{\pi^{(d-1)/2}}{\sqrt{2}\Gamma(d/2)}
\chi(1+\alpha)\nonumber\\
& & \times\left(\frac{d-1}{2}+\frac{3}{16}(d+8)(1-\alpha)\right).
\end{eqnarray}
Finally, the expressions of the (reduced) collision frequencies $\nu_1^*$, $\nu_2^*$, $\nu_4^*$, and $\nu_5^*$  can be found in the Appendix C of Ref.\ \cite{GF09} while $\nu_3^*$ and $\nu_6^*$ are given by \cite{note}
\begin{equation}
\label{a11} \nu_{3}^*=-\frac{\pi^{(d-1)/2}}
{d\Gamma\left(\frac{d}{2}\right)}\left(\frac{1+\omega}{2}\right)^{d-1}\chi_0
\frac{{\cal M}^2}{{\cal M}_{0}}(1+\alpha_0) \theta^{5/2}(1+\theta)^{-1/2},
\end{equation}
\begin{eqnarray}
\label{a12} \nu_{6}^*&=&-\frac{\pi^{(d-1)/2}}
{d(d+2)\Gamma\left(\frac{d}{2}\right)}\left(\frac{1+\omega}{2}\right)^{d-1}\chi_0
\frac{{\cal M}^2}{{\cal M}_{0}}(1+\alpha_0)\nonumber\\
& & \times \left(\frac{\theta}{1+\theta}\right)^{3/2}
\left[C+(d+2)(1+\theta)D\right],
\end{eqnarray}
where $C$ and $D$ are given by Eqs.\ (C7) and (C8), respectively, of Ref.\ \cite{GF09}.

\end{document}